\documentclass[12pt]{iopart}
\usepackage{epsfig,bm}

\begin{document}

\title{Meson-Meson Scattering in Relativistic Constraint Dynamics}
\author{Horace W. Crater$^{1}$ \& Cheuk-Yin Wong$^{2,3}$}

\address{$^{1}$University of Tennessee Space Institute, Tullahoma, TN
37388}

\address{ $^2$Physics Division, Oak Ridge National Laboratory, Oak Ridge, TN
37831}

\address{ $^3$Department of Physics, University of Tennessee, Knoxville, TN
37996}

\address{E-mail: hcrater@utsi.edu  \&  wongc@ornl.gov} 

\begin{abstract}
Dirac's relativistic constraint dynamics have been successfully applied to
obtain a covariant nonperturbative description of QED and QCD bound states.
We use this formalism to describe a microscopic theory of meson-meson
scattering as a relativistic generalization of the nonrelativistic
quark-interchange model developed by Barnes and Swanson.
\end{abstract}

In order to rule out false signals for the onset of the formation of a
quark-gluon plasma one needs a reliable relativistic meson-meson
scattering formalism. For example the dissociative process $\pi+
J/\psi \rightarrow D+{\bar D}^*$ could, by rapidly taking the $J/\psi
$ out of the picture, mimic the suppression of $J/\psi $ production
thought to occur in a quark-gluon plasma at high temperatures
\cite{Mat86}-\cite{Won04}. A nonrelativistic formalism for such a
process in the microscopic quark-interchange picture of meson-meson
scattering was developed by Barnes and Swanson \cite{barnes} and later
supplemented by a detailed quark model by Wong, Barnes, and Swanson
\cite{wong}. Here we show how to extend the quark-interchange model to
the relativistic domain using constraint dynamics, which has been
successfully applied to two-body bound state problems in QED
\cite{crater1}, QCD \cite{crater2}, and to two-body nucleon-nucleon
scattering \cite{crater3}.

The two-body relativistic wave equations of constraint dynamics can be
derived from the Bethe Salpeter Equation. They have their origins however in
classical relativistic mechanics where one starts with two mass shell
constraints and introduces interactions $\Phi_i$ (here world scalar
interactions) 
\begin{eqnarray}
\fl ~~~~~~\mathcal{H}_{i}^{0} &=&p_{i}^{2}+m_{i}^{2}\rightarrow
p_{i}^{2}+M_{i}^{2}\equiv \mathcal{H}_{i}\equiv p_{i}^{2}+m_{i}^{2}+\Phi
_{i}(x_{1}-x_{2},p_{1},p_{2});~~~~i=1,2,
\end{eqnarray}
in such a way that the constraints are compatible
$\{H_{1},H_{2}\}\approx 0$. These constraints in turn imply that the
interaction potentials satisfy a relativistic third-law condition
\begin{equation}
\Phi _{1}=\Phi _{2}=\Phi (x_{12\perp },p_{1},p_{2})\equiv \Phi _{w},
\end{equation}
and that they depend, as Fig.\ 1$a$ indicates, on its ``perped''
component $ x_{12\perp}=(x_1-x_2)_{\perp}$ perpendicular to the total
momentum $P=p_1+p_2 $. The relative time is covariantly eliminated
since in the CM system $r_{12} {\Large \equiv }\sqrt{x_{12\perp
}^{2}}{\Large =}\sqrt{\mathbf{r_{12}}^{2}} ;~t_{1}-t_{2}=0.$

\begin{figure}[h]
\hspace*{0.0cm} \includegraphics[angle=0,scale=0.75]{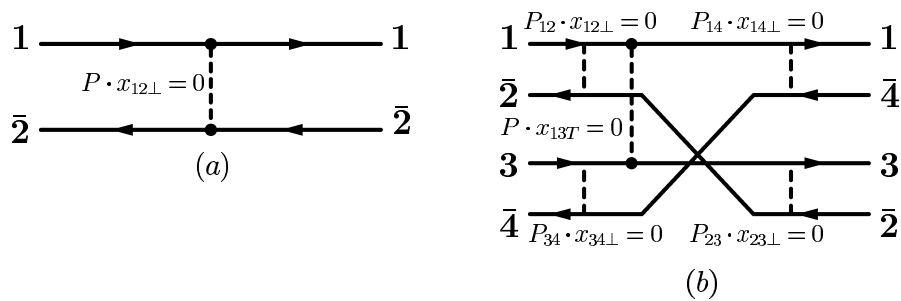} \vspace*{-16.0cm}
\caption{Diagram for ($a$) two-body bound state, and ($b$) meson-meson
scattering.}
\end{figure}

For two particles with spins, one has two Dirac equations \cite{crater4}
(here given for minimal scalar and vector interactions) instead of two
generalized mass shell constraints, 
\begin{equation}
\mathcal{S}_{i}\psi \equiv \gamma _{5i}(\gamma _{i}\cdot (p_{i}-\tilde{A}
_{i})+m_{i}+\tilde{S}_{i})\psi =0,~~~~i=1,2.
\end{equation}
Their compatibility ($[\mathcal{S}_{1},\mathcal{S}_{2}]\psi =0$) is
guaranteed if supersymmetry is added to the conditions that applied in
the two body spinless case. The vector and scalar interactions each
depend on underlying invariant functions $A(r)$ and $S(r)$. The
compatibility condition leads to an automatic incorporation of correct
spin-dependent recoil terms,

\begin{equation}
\fl ~~~~~~\tilde{A}_{i}^{\mu } =\tilde{A}_{i}^{\mu }(A(r),x_{\perp
},p_{1},p_{2},\gamma _{1},\gamma _{2});~\ \tilde{S}=\tilde{S}
_{1}(S(r),A(r),x_{\perp },p_{1},p_{2},\gamma _{1},\gamma _{2}).
\end{equation}
The two-body Dirac equations can be put into a simple and local 4-component
Schr\"{o}dinger-like form. In the case of lowest order QED, they have an
exact solution \cite{crater5} for singlet positronium that agrees with
standard perturbative results. Thus, they are less likely to produce
spurious results when applied to QCD.

Using such a formalism, we obtain very good results for the entire meson
spectrum from the light pion to the heavy upsilon states \cite{crater2}. The
nonperturbative structures in our equations provide for chiral symmetry in
the sense that the pion (although not its excited states or the $\rho $)
behave like a Goldstone boson.

The compatibility conditions for four spinless particles, $
\{H_{i},H_{j}\}\approx 0$, unlike their two body counterpart, are not
tractable as the set of two-body momenta are not separately conserved
and three- and four-body interactions are needed for full
compatability,
\begin{equation}
\fl \mathcal{H}_{i0}=p_{i}^{2}+m_{i}^{2}\rightarrow \mathcal{H}_{i}\equiv
p_{i}^{2}+m_{i}^{2}+\sum_{j\neq i}\Phi _{ij}(x_{ij\perp })+\sum_{j\neq k\neq
i}\Phi _{ijk}+\Phi _{1234}, i=1,2,3,4
\end{equation}

Previously, we used Dirac's constraint dynamics to obtain a Hamiltonian
formulation of the relativistic $N$-body problem in a separable two-body
basis in which the particles interact pair-wise through scalar and vector
interactions by neglecting the many-body interactions \cite{Won01}. The
resultant $N$-body Hamiltonian is relativistically covariant and can be
separated in terms of the center-of-mass and the relative motion of any
two-body subsystem. The two-body wave functions can be used as basis states
to evaluate reaction matrix elements in the general $N$-body problem. In
such a formalism, there is however the difficulty of determining the
commutation relations involving the creation or annihilation operators of
particles that belong to different composites.

Sazdjian \cite{saz1} has found alternatively that compatibility can be
obtained if one demands that the two-body interactions depend on the
component of the relative coordinates transverse to the total momentum
of the four-body system instead of the two-body system. In this
formalism, one introduces the transverse ($T$) component
\begin{eqnarray}
x_{ijT}=x_{ij}+ (x_{ij}\cdot P) \,P/\sqrt{
-P^{2}},
\end{eqnarray}
where $P=p_{1}+p_{2}+p_{3}+p_{4}$, $x_{ij}=x_{i}-x_{j}$, and one
assumes $\Phi _{ij}=\Phi _{ij}(x_{ijT})$. This formalism is suited for
bound systems. Below we shows its adaptation to the scattering
problem.

We review the nonrelativistic approach, starting with the orthogonality and
completeness conditions 
\begin{equation}
\langle \mathbf{p}_{1}^{\prime }\mathbf{,p}_{2}^{\prime }\mathbf{|p}
_{1}^{\prime \prime }\mathbf{,p}_{2}^{\prime \prime }\mathbf{\rangle }
\mathbf{=}\delta ^{3}\mathbf{(p}_{1}^{\prime }\mathbf{-p}_{1}^{\prime \prime
}\mathbf{)\delta }^{3}\mathbf{(p}_{2}^{\prime }\mathbf{-p}_{2}^{\prime
\prime }),
\end{equation}
\begin{equation}
~1_{12p}=\int d^{3}p_{1}^{\prime }d^{3}p_{2}^{\prime }|\mathbf{p}
_{1}^{\prime }\mathbf{,p}_{2}^{\prime }\mathbf{\rangle \langle p}
_{1}^{\prime }\mathbf{,p}_{2}^{\prime }|,
\end{equation}
so that with the wave function defined by $\langle \mathbf{p}_{1}^{\prime }
\mathbf{,p}_{2}^{\prime }|M(\mathbf{P})\rangle =\delta ^{3}(\mathbf{P}
_{12}^{\prime }\mathbf{-P}_{12})\tilde{\psi}_{P}(\mathbf{p}_{12}^{\prime }),$
the scalar products of meson wave functions is simply given by{\Large \ } 
\begin{equation}
\langle M(\mathbf{Q})|M(\mathbf{P})\rangle =\delta ^{3}(\mathbf{P}^{\prime }
\mathbf{-Q})\int d^{3}p_{12}\tilde{\psi}_{Q}^{\ast }(\mathbf{p}_{12}^{\prime
})\tilde{\psi}_{P}(\mathbf{p}_{12}^{\prime }).
\end{equation}
Using the above orthogonality, completeness conditions, and wave function,
we can construct the meson scattering amplitude for the reaction $\mathbf{P}
_{12}\mathbf{+P}_{34}\mathbf{\rightarrow Q}_{14}\mathbf{+Q}_{32}$ shown in
Fig.\ 1($b$) in terms of the momentum matrix elements of the interaction
potential, 
\begin{eqnarray}
&&\langle M(\mathbf{Q}_{14})M(\mathbf{Q}_{23});\mathbf{Q}|V(\mathbf{x}
_{13})|M(\mathbf{P}_{12})M(\mathbf{P}_{34});\mathbf{P}\rangle   \nonumber \\
&=&\int 
d^{3}q_{1}^{\prime }d^{3}q_{3}^{\prime }d^{3}p_{1}^{\prime
}d^{3}p_{3}^{\prime }
\langle \mathbf{q}_{1}^{\prime }\mathbf{,q}_{2}^{\prime }\mathbf{,q}
_{3}^{\prime }\mathbf{,q}_{4}^{\prime }|V(\mathbf{x}_{13})|\mathbf{p}
_{1}^{\prime }\mathbf{,p}_{2}^{\prime }\mathbf{,p}_{3}^{\prime }\mathbf{,p}
_{4}^{\prime }\rangle   \nonumber \\
&&\times \tilde{\psi}_{\mathbf{Q}_{14}}(\mathbf{q}_{14}^{\prime })\tilde{\psi
}_{\mathbf{Q}_{32}}(\mathbf{q}_{32}^{\prime })\tilde{\psi}_{\mathbf{P}_{12}}(
\mathbf{p}_{12}^{\prime })\tilde{\psi}_{\mathbf{P}_{34}}(\mathbf{p}
_{34}^{\prime })
\end{eqnarray}
The coordinate completeness condition then leads finally to the results of
Barnes, Swanson and Wong in \cite{barnes} and \cite{wong}, 
\begin{eqnarray}
\fl 
&&\langle M(\mathbf{Q}_{14})M(\mathbf{Q}_{23});\mathbf{Q}|V(\mathbf{x}
_{13})|M(\mathbf{P}_{12})M(\mathbf{P}_{34});\mathbf{P}\rangle   \nonumber \\
&\mathbf{=}&\mathbf{\delta }^{3}\mathbf{(Q}_{14}\mathbf{+Q}_{32}\mathbf{-P}
_{12}\mathbf{-P}_{34}\mathbf{)}\int d^{3}p_{1}^{\prime }d^{3}q_{1}^{\prime }
\tilde{V}\mathbf{(p}_{1}^{\prime }\mathbf{-q}_{1}^{\prime }\mathbf{)}\tilde{
\psi}_{\mathbf{Q}_{14}}\mathbf{(q}_{1}^{\prime }\mathbf{-}\frac{m_{1}}{M_{14}
}\mathbf{Q}_{14}\mathbf{)}  \nonumber \\
&&\mathbf{\times }\tilde{\psi}_{\mathbf{Q}_{32}}\mathbf{(}\frac{m_{2}}{M_{32}
}\mathbf{Q}_{32}\mathbf{-P}_{12}\mathbf{+p}_{1}^{\prime }\mathbf{)}\tilde{
\psi}_{\mathbf{P}_{12}}\mathbf{(p}_{1}^{\prime }\mathbf{-}\frac{m_{1}}{M_{12}
}\mathbf{P}_{12}\mathbf{)}\tilde{\psi}_{\mathbf{P}_{34}}\mathbf{(q}
_{1}^{\prime }\mathbf{-Q}_{14}\mathbf{+}\frac{m_{4}}{M_{34}}\mathbf{P}_{34}
\mathbf{)}.  \label{a}
\end{eqnarray}

We now display an analogous formalism in the relativistic case in which, as
in the nonrelativistic case, the key ingredients are the scalar product,
orthogonality, and completeness conditions.

Scalar products \cite{saz2} in the relativistic case are complicated by the
fact that the relativistic effective potentials are energy dependent (as
occurs for example in the one-body Klein-Gordon equation). In a general case
this may lead to important contributions but in the Born approximation that
we follow here it can be ignored.

We introduce here the completeness and orthogonality conditions, 
\begin{eqnarray}
1_{12p} &=&\int d^{4}p_{12}^{\prime }d^{4}P_{12}^{\prime }|p_{1}^{\prime
},p_{2}^{\prime }\rangle \delta (P_{12}^{\prime }\cdot \hat{P}+w)\delta
(p^{\prime }\cdot \hat{P})\langle p_{1}^{\prime },p_{2}^{\prime }|, 
\nonumber \\
\langle p_{1}^{\prime \prime },p_{2}^{\prime \prime }|p_{1}^{\prime
},p_{2}^{\prime }\rangle  &=&\tilde{\delta}^{4}(p_{1}^{\prime \prime
}-p_{1}^{\prime },\tau )\tilde{\delta}^{4}(p_{2}^{\prime \prime
}-p_{1}^{\prime },\tau ),  \nonumber \\
\tilde{\delta}^{4}(p_{i}^{\prime \prime }-p_{i}^{\prime },\tau ) &\equiv
&\int dr_{i}\delta ^{4}(p_{i}^{\prime \prime }-p_{i}^{\prime }+r\hat{P}
_{12})\exp (-ir_{i}\tau {\Large )},
\end{eqnarray}
where $p^{\prime }=(\varepsilon _{2}p_{1}^{\prime }-\varepsilon
_{1}p_{2}^{\prime })/w$, and $P_{12}^{\prime }=p_{1}^{\prime
}+p_{2}^{\prime }$. Unlike the nonrelativistic case there is here a
$\hat{P},$ showing the dependence on the particular two-body state.

Using the wave function ($\hat{n}$ an arbitrary time-like unit vector) 
\[
\langle x_{1}^{\prime },x_{2}^{\prime }|M(\hat{P}_{12})\rangle \equiv \sqrt{
\frac{\left( {\normalsize -}\hat{P}_{12}{\normalsize \cdot }\hat{n}\right) }{
(2\pi )^{3}}}\exp i(P_{12}\cdot X_{12}^{\prime })\psi _{P_{12}}(x_{12\perp
}^{\prime })
\]
and completeness conditions, we obtain the scalar product
\begin{eqnarray}
& &\langle M(Q_{12}|M(P_{12})\rangle 
\nonumber\\
&=&\tilde{\delta}^{4}(P_{12}-Q_{12},\tau
)(-\hat{P}_{12}\cdot \hat{n})\int d^{4}x_{12}^{\prime }\delta
(x_{12}^{\prime }\cdot \hat{P}_{12})\psi _{Q_{12}}^{\ast }(x_{12\perp
}^{\prime })\psi _{P_{12}}(x_{12\perp }^{\prime }).
\end{eqnarray}
The derivation of the meson-meson scattering amplitude parallels its
nonrelativistic counterpart until one gets to the the momentum space
matrix element of the potential $\langle q_{1}^{\prime
},q_{2}^{\prime },q_{3}^{\prime },q_{4}^{\prime }|\Phi (x_{13
T})|p_{1}^{\prime },p_{2}^{\prime },p_{3}^{\prime },p_{4}^{\prime
}\rangle $ that is analogue of the nonrelativistic matrix element
$\mathbf{\langle q}_{1}^{\prime } \mathbf{,q}_{2}^{\prime
}\mathbf{,q}_{3}^{\prime }\mathbf{,q}_{4}^{\prime }
\mathbf{|}V\mathbf{(x}_{13}\mathbf{)|p}_{1}^{\prime
}\mathbf{,p}_{2}^{\prime }\mathbf{,p}_{3}^{\prime
}\mathbf{,p}_{4}^{\prime }\mathbf{\rangle }$.  The problem is that the
bra and ket momentum states in the relativistic expression belong to
different mesons.

The initial state orthogonality condition below (final state condition
is similar) shows the explicit dependence on meson momenta
$P_{12}$ and $P_{34}$,
\begin{eqnarray}
& & \!\!\!\!
\langle p_{1}^{\prime \prime },p_{2}^{\prime \prime
},p_{3}^{\prime \prime },p_{4}^{\prime \prime };\hat{P}_{12},\hat{P}
_{34}|p_{1}^{\prime },p_{2}^{\prime },p_{3}^{\prime },p_{4}^{\prime };\hat{P}
_{12},\hat{P}_{34}\rangle   \nonumber \\
& &\!\!\!\! 
=\int dr_{1}dr_{2}dr_{3}dr_{4}\delta ^{4}(p_{1}^{\prime \prime
}-p_{1}^{\prime }+r_{1}\hat{P}_{12})\delta ^{4}(p_{2}^{\prime \prime
}-p_{2}^{\prime }+r_{2}\hat{P}_{12})  \nonumber \\
& & \!\!\!\!\times 
\delta ^{4}(p_{3}^{\prime \prime }-p_{3}^{\prime }+r_{3}\hat{P}
_{34})\delta ^{4}(p_{4}^{\prime \prime }-p_{4}^{\prime }+r_{4}\hat{P}
_{34})\exp (-i(r_{1}+r_{2}+r_{3}+r_{4})\tau ).
\end{eqnarray}
Our postulate for different sets of mesons in the bra and ket states is one
that uses the total four momentum unit vector $\hat{n}$ of the four quark
system in place of the constituent four momenta,
\begin{eqnarray}
&&\!\!\!\!\!\!\!\!\!\!\!
\langle q_{1}^{\prime },q_{2}^{\prime },q_{3}^{\prime },q_{4}^{\prime
}|p_{1}^{\prime },p_{2}^{\prime },p_{3}^{\prime },p_{4}^{\prime }\rangle
\equiv \langle q_{1}^{\prime },q_{2}^{\prime },q_{3}^{\prime },q_{4}^{\prime
};\hat{Q}_{14},\hat{Q}_{32}|p_{1}^{\prime },p_{2}^{\prime },p_{3}^{\prime
},p_{4}^{\prime };\hat{P}_{12},\hat{P}_{34}\rangle   \nonumber \\
&=&\int dr_{1}dr_{2}dr_{3}dr_{4}\delta ^{4}(q_{1}^{\prime }-p_{1}^{\prime
}+r_{1}\hat{n})\delta ^{4}(q_{3}^{\prime }-p_{3}^{\prime }+r_{3}\hat{n}) 
\nonumber \\
&&\times \delta ^{4}(q_{2}^{\prime }-p_{2}^{\prime }+r_{2}\hat{n})\delta
^{4}(q_{4}^{\prime }-p_{4}^{\prime }+r_{4}\hat{n})\exp
(-i(r_{1}+r_{2}+r_{3}+r_{4})\tau ).
\end{eqnarray}

The physical assumption this reflects is that in the collision process the
individual mesons lose their identity and momentarily we have a four body
system as described by the Sazdjian formalism. In a like manner the momentum
matrix element of the potential 
\begin{eqnarray}
&&\langle q_{1}^{\prime },q_{2}^{\prime },q_{3}^{\prime },q_{4}^{\prime
}|\Phi (x_{13 T})|p_{1}^{\prime },p_{2}^{\prime },p_{3}^{\prime
},p_{4}^{\prime }\rangle   \nonumber \\
&=&\langle q_{1}^{\prime },q_{2}^{\prime },q_{3}^{\prime },q_{4}^{\prime
}|\Phi (x_{13}\cdot (1+\hat{n}\hat{n}))|p_{1}^{\prime },p_{2}^{\prime
},p_{3}^{\prime },p_{4}^{\prime }\rangle 
\end{eqnarray}
is one that reflects the Sazdjian hypothesis of coordinate dependence
only through its component perpendicular to the total momentum. We
thus have a hybrid model in which the meson wave functions have the
usual two-body perped variable ($\perp$) dependence but the potential
has the four-body transversality ($T$) dependence. The diagram in
Fig.\ (1b) details the hybrid nature of the two combined constraint
formalisms.

We obtain finally an expression that is a relatively simple
three-dimensional but covariant generalization of the nonrelativistic
expression given in Eq.(\ref{a}) 
\begin{eqnarray}
&&\langle M(Q_{14})M(Q_{23});Q|\Phi (x_{13 T})
|M(P_{12})M(P_{34});P\rangle   \nonumber \\
&{\Large =}&\sqrt{\frac{\hat{n}\cdot \hat{P}_{12}\hat{n}\cdot \hat{Q}_{14}}{
\hat{n}\cdot \hat{P}_{34}\hat{n}\cdot \hat{Q}_{32}}}\tilde{\delta}
^{4}(P-Q,\tau )\int \delta (\hat{P}_{12}\cdot p_{12}^{\prime })\delta
(q_{14}^{\prime }\cdot \hat{Q}_{14})  \nonumber \\
&&\times \tilde{\psi}_{Q_{14}}(q_{14}^{\prime })\tilde{\psi}
_{Q_{32}}(q_{32}^{\prime })\tilde{\psi}_{P_{12}}(p_{12}^{\prime })\tilde{\psi
}_{P_{34}}(p_{34}^{\prime })\Phi \lbrack (p_{13}^{\prime }-q_{13}^{\prime
})_{T}]d^{4}q_{1}^{\prime }d^{4}p_{1}^{\prime }.
\end{eqnarray}
Our aim now is to apply the relativistic quark model wave functions we have
developed to compute meson-meson cross sections.

\vspace*{0.5cm}
\noindent \textbf{Acknowledgment}

\vspace*{0.5cm}
\noindent The authors would like to thank T. Barnes and E. S. Swanson for
helpful discussions. This research was supported by the NSF under Contract
No. NSF-PHY-0244819.

\end{document}